# SOP Sensing in Hollow-Core Fiber: Forward Sensing using COTS Transponder and Localization with Coherent OTDR

Adrish Sahu(1*), Rajiv Boddeda(1), Arnaud Dupas(1), Carina Castineiras Carrero(1), Haïk Mardoyan(1), Sterenn Guerrier(1), Matteo Lonardi(2), Javier Errea Moreno(3), Peng Li(4), Shuhai Li(4), Lei Zhang(4), Jie Luo(4), Daniel Peterson(5), and Jérémie Renaudier(1)

(1) Nokia Bell Labs, Massy, France; *adrish.sahu@nokia.com
(2) Nokia, Vimercate, Italy.
(3) Nokia Networks France, Optics division, Route de Villejust, Nozay, France
(4) State Key Laboratory of Optical Fiber and Cable Manufacture Technology, Yangtze Optical Fibre and Cable (YOFC), Wuhan, China.
(5) Nokia, Richardson, Texas, USA.

**Abstract** *We experimentally investigate and compare distributed state-of-polarization sensing in support-tube hollow-core and standard single-mode fiber. Despite significantly weaker Rayleigh backscattering and reduced polarization sensitivity in hollow-core fiber, polarization perturbations remain observable and spatially localizable, demonstrating the feasibility of SOP-based distributed sensing in hollow-core fiber. ©2026 The Author(s)*

## Introduction

The state of polarization (SOP) of light in optical fibers serves as a valuable tool for physical layer monitoring, as environmental or mechanical disturbances induce detectable polarization variations along the optical link. Prior demonstrations on standard single-mode fiber (SSMF) show that SOP can reveal stress, vibration, or cable motion, though interpretation is complicated by strong random birefringence and accumulated polarization mode coupling along the fiber [1-3]. Consequently, SOP evolution after propagation along the SMF often resembles a noisy random walk, making weak events difficult to detect and localize [2,3]. By using coherent distributed acoustic sensing, one can localize the polarization and reconstruct the polarization evolution as a function of distance. Hollow core fiber has enabled low-latency and ultralow-nonlinearity optical transmission with substantially reduced light–matter interaction [4-6]. Compared to SSMF, HCF exhibits weaker refractive-index perturbations [7,8], distinct birefringence characteristics, and reduced scattering. In this article, we study this evolution for both standard SMF and in Support Tube (ST-HCF) by creating a controlled perturbation along the fiber.

Additionally, the polarization-based sensing in HCF, particularly in support-tube architectures, remains largely unexplored. It is not yet known whether disturbances applied along an HCF span produce stronger, weaker, or simply qualitatively different SOP signatures compared with SMF. This represents a gap in the state of the art: the community lacks a systematic comparison of distributed SOP sensing in HCF versus SMF under controlled perturbation.

In this paper, using advanced distributed sensing techniques and polarization sensing based on coherent digital signal processing available in commercial coherent transceivers, we characterize the polarisation sensibility of ST-HCF compared to SSMF. In addition, we show that one can measure and localize a polarization perturbation event on a ST-HCF using frequency stabilized coherent optical time domain reflectometry (OTDR). This demonstrates the capability to precisely identify and characterize external influences affecting the polarization state within HCFs, despite the inherent robustness.

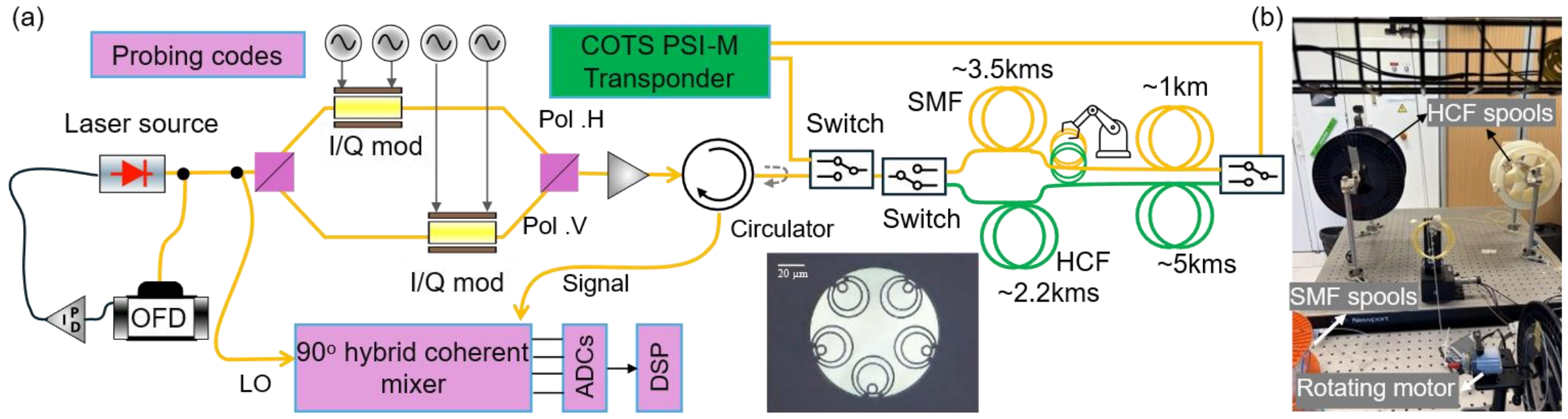

**Fig. 1:** (a) Setup of the experiment with COTS PSI-M for forward polarization sensing and Fiber sensing setup. (inset) cross-section of the ST-HCF and (b) Robotic arm made using a motor to create perturbations on the HCF and the SSMF fiber.

## Experimental setup

Continuous probing coherent coded OTDR is a powerful distributed sensing technique that allows characterization of optical fiber properties along the full length of the fiber [9,10]. Unlike conventional OTDR, which is based on time-of-flight measurements of backscattered light, the use of continuous probing codes enables higher spatial resolution and faster acquisition [10]. As the probing signal is continuous, more energy can be launched within one probing interval (equal to the total fiber propagation time), allowing for faster extraction of the correlation signal. Additive noise is more critical due to the weak backscattered signal, especially in HCF, together with speckle noise along the fiber. Hence, we employ a stabilized laser source using a custom compact Optical Frequency Discriminator to lower the laser frequency noise in the frequency range of interest [10]. In Fig. 1a, we show the coherent OTDR setup where a circulator is used to recover the Rayleigh scattering from the fiber. We use a switch after the circulator to switch to forward sensing where coherent signal generated and decoded by a commercial off the shelf transponder (COTS), consistent with prior polarization-based sensing approaches using coherent receivers [11]. By using the readily functionalities available in the Nokia PSI-M COTS transponder, we recover the variation of the polarization as a function of time with a speed faster than 100 Hz. We then switch between either 7.2 km ST-HCF (inset shows the cross-section of the fiber) or 4.5 km SSMF fiber to probe the polarization variations. At the end of both fibers, a switch closes the loop to the transponder. The perturbations are created by a robotic arm located at ~3.5 km and ~2.2 km from the beginning which moves in a periodic fashion. In Fig.1b, we show the rotating motor with a pulley which serves as a moving arm connected to a set of spools for both SSMF and HCF.

## Forward sensing in HCF and SSMF

When a fiber is subjected to rotational strain, the induced circular birefringence causes a rotation of the polarization plane of linearly polarized light during propagation. The amount of rotation is proportional to the applied twist rate and the fiber's photoelastic coefficients [12]. This effect forms the basis for the polarization response of a fiber under controlled twisting which is used in the design of polarisation controllers.

We create a periodic rotation at approximately 2 Hz and then study the response of both fibers by switching between HCF and SSMF using the forward sensing setup. In Fig. 2a, we show the three stokes parameter of the received signal after propagating through the fiber. The top graph shows the measured stokes variations for SSMF fiber with a clear signature of saturation while the bottom graph for the HCF shows a clean sinusoidal variation of the polarisation. In Fig. 2b, we show the evolution of SOP on the Poincaré sphere. In Fig. 2c we show the maximum geodesic displacement from the initial polarisation state. Our measurements show that for the same amount of rotational strain, the polarization rotation observed in HCF is at least 2.3 times smaller compared to SMF. This reduction in sensitivity is attributed to the fundamental difference in light guidance: In SMFs, light propagates entirely within the solid silica core, which is highly susceptible to the photoelastic effect. The stress induced by twisting directly alters the refractive index tensor of the silica, leading to significant circular birefringence. In HCFs, the light propagates predominantly in the air core. Air does not exhibit significant photoelastic effect. While the silica cladding structure experiences strain, the interaction of the guided mode with this strained silica is minimal. The optical field is largely decoupled from the solid material that would otherwise induce strong photoelastic effects. The unique geometry of HCFs, particularly those with anti-resonant structures, can also contribute to a more stable polarization state by minimizing the effective stress-optic coefficient experienced by the guided mode.

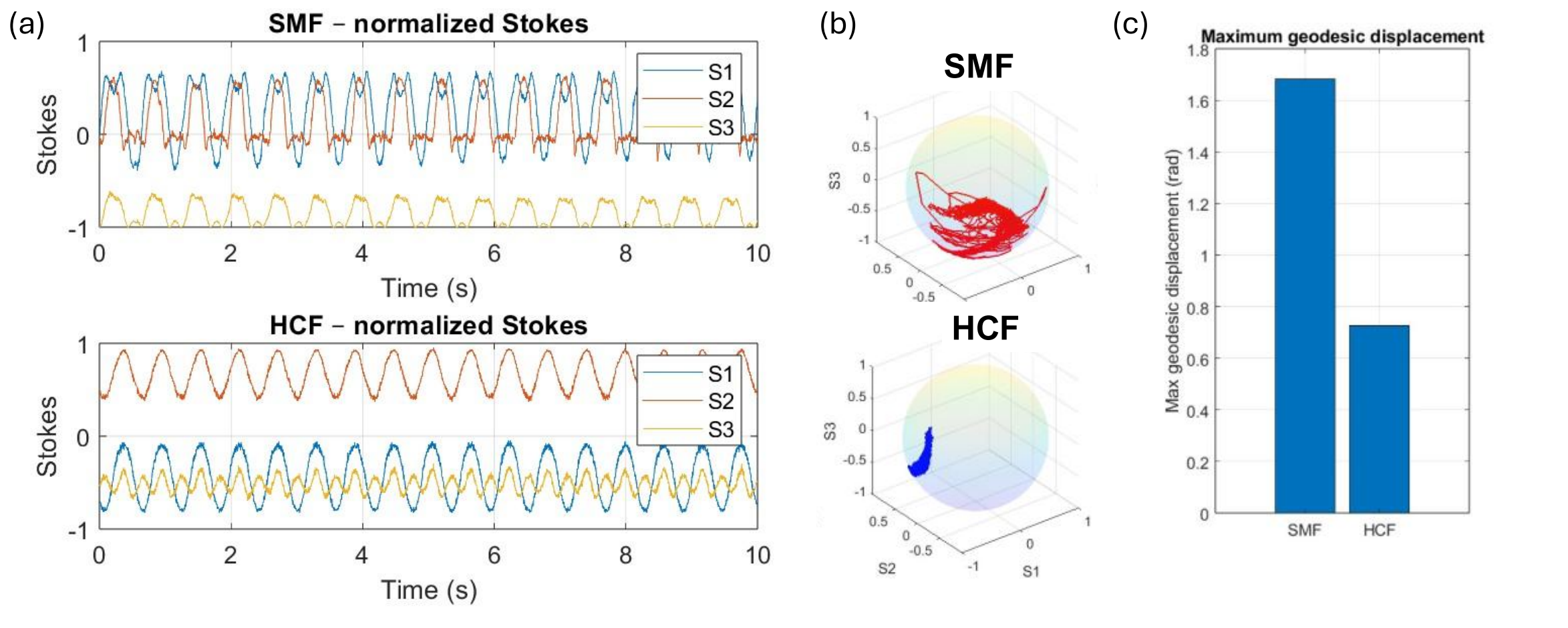

**Fig. 2:** Forward sensing polarization analysis in SMF and HCF. (a) Time-domain normalized Stokes parameters, (b) Corresponding Poincaré-sphere trajectories, and (c) Maximum geodesic displacement is shown for both SMF and HCF.

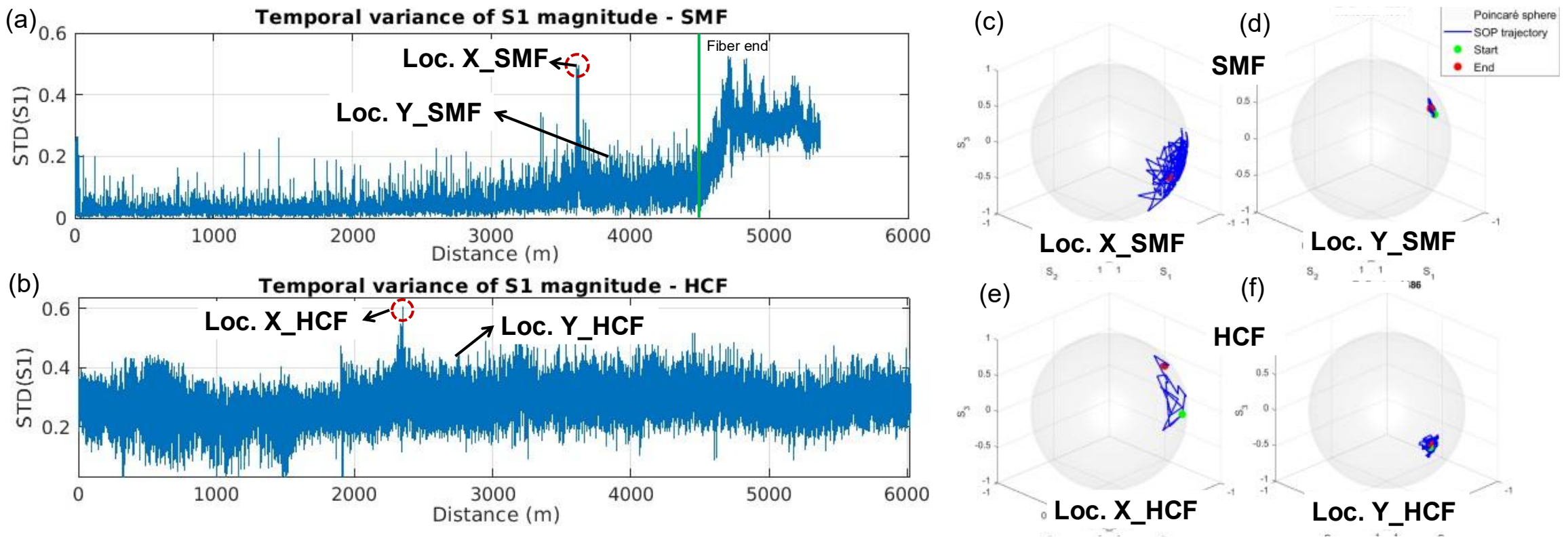

**Fig. 3:** Spatial localization of a known event in SMF (a), and HCF (b). Evolution of polarization on the Poincaré sphere in SMF (c,d) and HCF (e,f), measured at the reflector X SMF, Y SMF and X HCF, Y HCF respectively.

This reduced sensitivity to rotational strain results in a state-of-polarization evolution in HCF characterized by smaller deviations from the initial polarization state compared with SMF.

### Spatial Localization via Stokes Parameters

We switch to the coherent OTDR setup to analyze Rayleigh backscattering coming from both the fibers. We use probing code design derived from [9], thanks to the stabilized laser source we can use code lengths reaching ~ few milliseconds and acquisition time of few seconds. The code is made of $2^{21}$ symbols at up to 500 Mbaud, leading to a 0.3 m native gauge length, and the permanent repetition of the code allows to capture mechanical events over a 120 Hz bandwidth. By analyzing the Rayleigh reflected signal from each segment, we estimate the Jones Matrix vector, which is then used to derive the Stokes parameters from this element [13].

Unlike forward sensing where the polarization fluctuations are cumulated along the fiber, here we focus solely on estimating variations associated with an isolated segment, which introduces greater uncertainty in the estimation of the Stokes vector. To enable spatial localization along the fiber, the temporal variation of the $S_1$ Stokes parameter was therefore evaluated at each reflector position. We then estimate the standard deviation of $S_1$ as a function of distance which allows us to isolate the location of the perturbation.

In Fig. 3 we show the standard deviation of S1 of the light reflected all along the fiber from SMF (a) and HCF (b). We select from the trace two reflectors located at location X_SMF and Y_SMF shown in Fig. 3a where X_SMF corresponds to the location of perturbation. We show in the Fig. 3c,d the trajectories correspond to the temporal evolution of the normalized Stokes parameters extracted at X_SMF and Y_SMF. In Fig. 3b, we select the dominant reflector X_HCF corresponding to the location of perturbation and Y_HCF just after the perturbation. In both cases, X positions are chosen at the location of perturbation and Y is just after the perturbation. We see the evolution of the SOP represented on the Poincaré sphere for the SMF and HCF clearly showing a larger variation in X compared to Y due to the fiber strain by the motor. In Fig. 3b, we observe that the standard deviation is larger in HCF due to the intrinsically weaker Rayleigh backscattering in hollow-core fiber. Nevertheless, this can be corrected by increasing the launched power in the fiber. The polarization perturbation remains clearly visible in both fibers. Compared to SMF, the overall magnitude of polarization evolution in HCF is reduced, indicating a weaker sensitivity of the guided mode to mechanically induced birefringence. A clear localized feature remains visible at the perturbation position, allowing spatial identification of the event. Importantly, the perturbation signature in HCF remains distinguishable from the noise floor, demonstrating the effectiveness of the variance-based approach even with noisy conditions.

## 6. Conclusion

We measured the polarization sensitivity of HCF compared to SSMF using forward sensing setup with a COTS transponder. ST-HCF is shown to exhibit smaller polarization fluctuations under identical applied rotational strain compared to SSMF. Phase-sensitive OTDR confirmed this sensitivity difference between HCF and SSMF while providing spatially resolved identification of perturbations.

## Acknowledgements

This work has been supported by the EU's Horizon Europe, under grant agreement No 101189545 – SENSEI.